\documentclass[aps, twocolumn, superscriptaddress, showpacs,
citeautoscript, floatfix]{revtex4-1}
\usepackage{array}
\usepackage{graphicx}
\usepackage{bm}
\newcommand* {\vek}[1]{{\bm{\mathrm{#1}}}}
\newcommand* {\kk}{\vek{k}}
\newcommand* {\qq}{\vek{q}}
\newcommand* {\frack}[2]{{\Ts\frac{#1}{#2}}}
\newcommand* {\Ds}{\displaystyle}
\newcommand* {\Ts}{\textstyle}
\begin{document}

\title{Effective Hamiltonian for protected edge states in graphene}

\author{H. Deshpande}
\affiliation{Department of Physics, Northern Illinois University,
DeKalb, Illinois 60115, USA}
\author{R. Winkler}
\affiliation{Department of Physics, Northern Illinois University,
DeKalb, Illinois 60115, USA}
\affiliation{Materials Science Division, Argonne National
Laboratory, Argonne, Illinois 60439, USA}

\date{March 28, 2017}

\begin{abstract}
  Edge states in topological insulators (TIs) disperse symmetrically
  about one of the time-reversal invariant momenta $\Lambda$ in the
  Brillouin zone (BZ) with protected degeneracies at
  $\Lambda$. Commonly TIs are distinguished from trivial insulators
  by the values of one or multiple topological invariants that
  require an analysis of the bulk band structure across the BZ. We
  propose an effective two-band Hamiltonian for the electronic
  states in graphene based on a Taylor expansion of the
  tight-binding Hamiltonian about the time-reversal invariant $M$
  point at the edge of the BZ.  This Hamiltonian provides a faithful
  description of the protected edge states for both zigzag and
  armchair ribbons though the concept of a BZ is not part of such an
  effective model.  We show that the edge states are determined by a
  band inversion in both reciprocal and real space, which allows one
  to select $\Lambda$ for the edge states without affecting the bulk
  spectrum.
\end{abstract}

\maketitle

A topological insulator (TI) is an insulator in the bulk with
topologically protected edge states that cross the gap so that the
edges are conducting.  This concept was first introduced by Kane and
Mele using a simple tight-binding (TB) model for the band structure
of graphene \cite{kan05, kan05a}.  Since then a wide range of
materials with these properties have been identified in two and
three dimensions (2D and 3D) \cite{has10, qi11}.  TIs can be
distinguished from trivial insulators without topological edge
states by the values of one or multiple topological invariants that
require an analysis of the bulk band structure across the Brillouin
zone (BZ).  In that sense TIs are considered conceptually different
from other problems in solid state physics that permit a description
local in $\kk$ space.

The first experimental verification of topologically protected edge
states was achieved for HgTe/CdTe quantum wells (QWs) \cite{koe07}
following a theoretical proposal by Bernevig, Hughes and Zhang
\cite{ber06a} based on a simple effective Hamiltonian, today known
as BHZ model.  Since then the BHZ model has been used in a wide
range of studies.  Liu et al.\ showed \cite{liu08} that it also
describes the edge states in InAs/GaSb QWs.  Zhou et al.\
demonstrated \cite{zho08} that the BHZ model can be solved exactly,
yielding analytical expressions for the edge states in HgTe/CdTe
QWs, see also Ref.~\onlinecite{son10c}.  We do not question the deep
insights that have emerged from the classification of solids based
on topological invariants.  But Zhou's work \cite{zho08} raises the
question to what extent TIs permit a description local in $\kk$
space \cite{effective}.  Is the concept of a BZ a necessary
prerequisite for protected edge states in a TI?  Graphene with its
simple TB description \cite{wal47} has served as an archetype for
TIs \cite{kan05, kan05a, hal88}, despite the fact that its intrinsic
SOC has been found to be small \cite{net09}.
We show here that a Taylor expansion of the graphene TB model about
the time-reversal invariant $\vek{M}$ point of the BZ (with
$\vek{M} \equiv - \vek{M}$) yields an effective Hamiltonian that
provides a faithful description local in $\kk$ space of the
protected edge states in both zigzag and armchair graphene ribbons.
While the proposed model is quite different from the more familiar
BHZ Hamiltonian, these models share a range of conceptual features,
some of which previously unrecognized, which suggests that these
features are common among TIs.  Quite generally \cite{kan05} the
edge states in TIs disperse symmetrically about one of the
time-reversal invariant momenta (TRIM) $\Lambda$ with protected
degeneracies at $\Lambda$.  We show that this $\Lambda$ is determined
by a band inversion in both reciprocal and real space, which allows
one to select $\Lambda$ without affecting the bulk spectrum.

\begin{figure}
  \includegraphics[width=0.9\columnwidth]{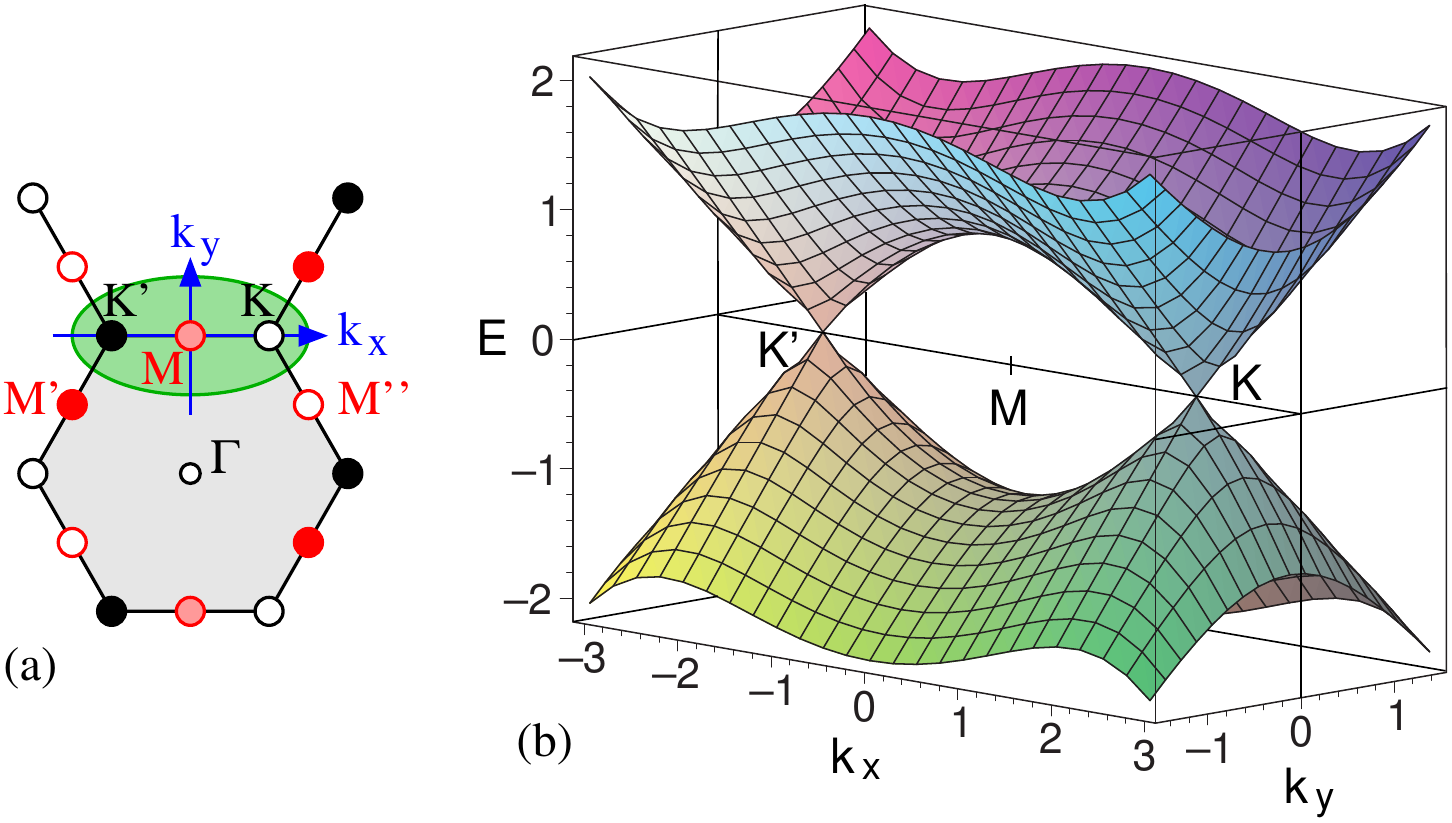}
  \caption{\label{fig:surface}(Color online) (a) Bulk BZ
  of graphene.  The region captured by the effective Hamiltonian
  (\ref{eq:ham-m}) is marked in green.  (b) Bulk band structure
  $E(\kk)$ of the Hamiltonian (\ref{eq:ham-m}) in the limit
  $\lambda_v = \lambda_i = \lambda_r = 0$.}
\end{figure}

In the following our conventions for the TB Hamiltonian follow
Refs.~\onlinecite{kan05, kan05a}, see also
Ref.~\onlinecite{prefac}\nocite{win10a}.  While the graphene BZ has two
inequivalent points $\vek{K}$ and $\vek{K}'$ (with
$\vek{K} \not\equiv - \vek{K}$), we have three inequivalent points
$\vek{M}$, $\vek{M}'$ and $\vek{M}''$, see
Fig.~\ref{fig:surface}(a).  Expanding the TB Hamiltonian for the
graphene $\pi$ bonds about $\vek{M} = (0, 2\pi/\sqrt{3})$, the
effective Hamiltonian up to second order in $\kk = (k_x, k_y)$
becomes
\begin{equation}
  \label{eq:ham-m}
  \arraycolsep 0.1em \hspace*{-0.8em}
  \begin{array}[b]{@{}>{\Ds}l@{}}
    H_M (\kk) =
    \left[\left(1 - \frack{1}{4} k_x^2 + \frack{1}{12} k_y^2 \right) \sigma_z
    + \frack{2}{\sqrt{3}} k_y \sigma_y \right] t
    - \sigma_x \lambda_v
    \\[0.5ex] \hspace{1em} {}
    - 4 k_x s_z \sigma_x \lambda_i
    + \bigl[ - \frack{2}{\sqrt{3}} s_x \sigma_y
    - \bigl(\frack{1}{2\sqrt{3}} k_y s_x
            + \frack{\sqrt{3}}{2} k_x s_y\bigr) \sigma_z
    \hspace*{-1.5em} \\[1.2ex] \hspace{1em} {}
    + \bigl(\frack{1}{8} k_x^2 s_x
            -\frack{1}{4} k_x k_y s_y
            +\frack{5}{24} k_y^2 s_x \bigr) \sigma_y\bigr] \lambda_r ,
  \end{array}
\end{equation}
where $s_i$ denotes spin operators and $\sigma_i$ are Pauli
matrices.  The first term describes the orbital motion characterized
by the nearest-neighbor hopping parameter $t$ (in the following
$t \equiv 1$).  The second term describes a staggered sublattice
potential weighted by $\lambda_v$ \cite{kan05, kan05a}. The third
term gives the intrinsic spin-orbit coupling (SOC) proportional to
$\lambda_i$.  The fourth term describes the Rashba SOC weighted by
$\lambda_r$.
In the following, $\qq$ refers to wave vectors in the BZ whereas
$\kk$ denotes wave vectors relative to the expansion point
$\qq_0$ of the effective Hamiltonian, i.e.,
$\qq = \qq_0 + \kk$.

\begin{figure}
  \includegraphics[width=\columnwidth]{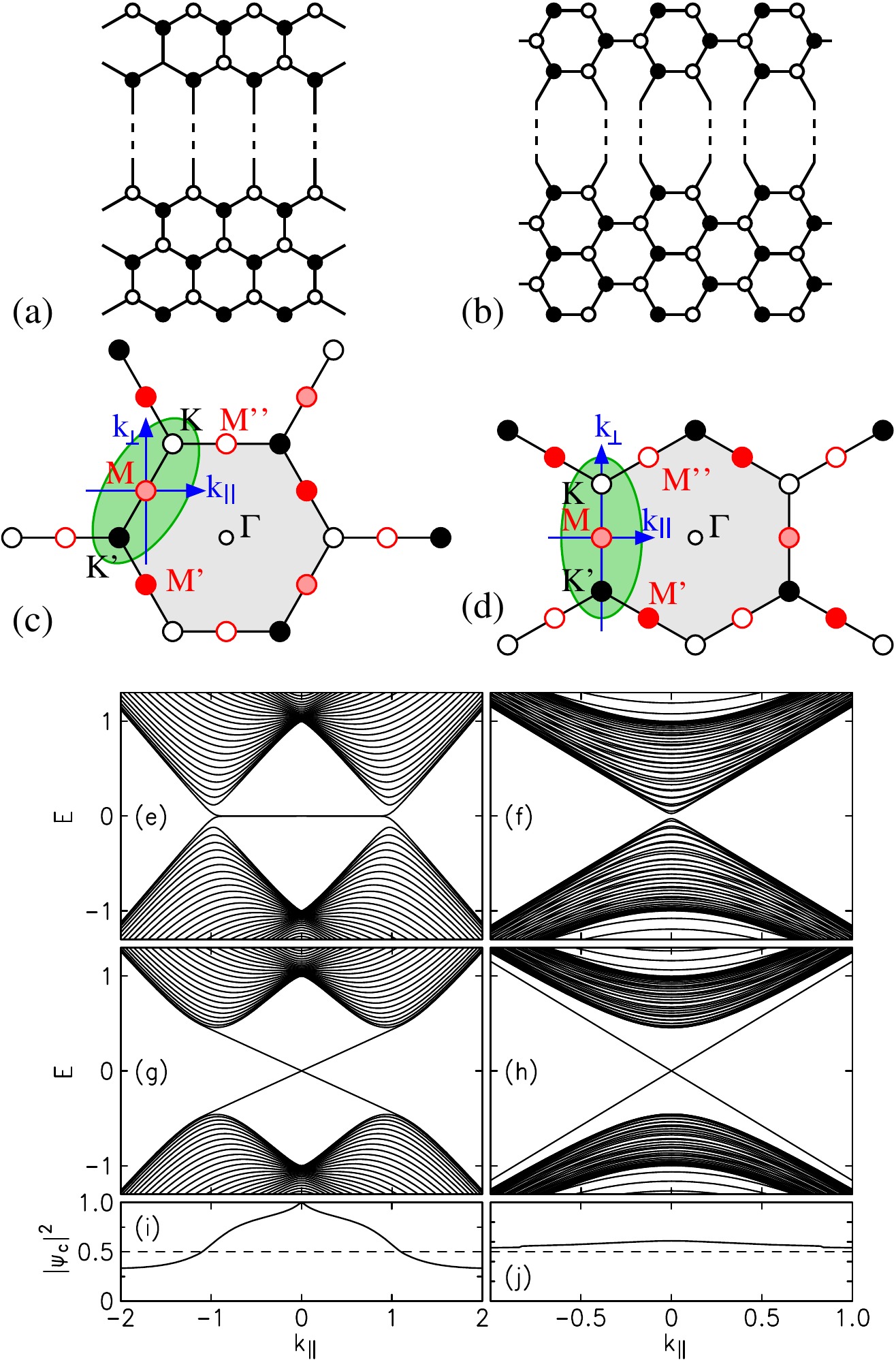}
  \caption{\label{fig:ribbon}(Color online) Crystal structure of a
  graphene ribbon with (a) zigzag and (b) armchair edges.  Bulk BZ
  of graphene corresponding to (c) zigzag and (d) armchair ribbons.
  The region of the BZ captured by the Hamiltonian (\ref{eq:ham-m})
  is marked in green.  Band structure $E(k_\|)$ of (e) zigzag and
  (f) armchair ribbons in the absence of SOC and (g), (h) for
  $\lambda_i = 0.2$.  (i), (j) Squared magnitude of the up component
  of pseudospin $\sigma_z$  of the lowest $E>0$ bulk eigenstates.
  The width of the ribbons is $w=40$ and we used $b_e = -10$.}
\end{figure}

First we discuss the properties of $H_M$ in the absence of SOC.
Unlike the Bloch states at the $K$ point \cite{nov05a}, the Bloch
states at the $M$ point are nonzero on both sublattices of the
graphene structure so that here $\sigma_z$ does not permit an
interpretation as sublattice pseudospin.  For $\lambda_v = 0$, the
dispersion becomes [Fig.~\ref{fig:surface}(b)]
\begin{equation}
  E_\pm (\kk) = \pm \sqrt{(1 - \frack{1}{4}k_x^2)^2
   + k_y^2 (\frack{3}{2} - \frack{1}{24} k_x^2) + \frack{1}{144} k_y^4} ,
\end{equation}
where the upper (lower) sign corresponds to the conduction (valence)
band.  For these bands the $M$ point is a saddle point.  For
$k_y = 0$, the dispersion becomes
$E_\pm (k_x,0) = \pm (1 - \frack{1}{4} k_x^2)$ so that the bands
touch at the points $(\pm 2,0)$, which mimics the dispersion near
the points $K$ and $K'$ of the BZ, the precise coordinates of which
are $(\pm 2\pi/3,0)$.  Indeed, if we substitute
$k_x \rightarrow k_x \pm 2$, the Hamiltonian (\ref{eq:ham-m}) is
unitarily equivalent to (ignoring Rashba SOC for simplicity)
\begin{equation}
  \label{eq:ham-k}
  \arraycolsep 0.15em
  \begin{array}[b]{rl}
    H_K (\kk) = & \pm k_x \sigma_x - \frack{2}{\sqrt{3}} k_y \sigma_y
    + (\frack{1}{4} k_x^2 - \frack{1}{12} k_y^2) \sigma_x
    \\[1.5ex] &
    + \lambda_v \sigma_z \pm 4 (2 \pm k_x) s_z \lambda_i \sigma_z .
  \end{array}
\end{equation}
For small $\kk$, Eq.\ (\ref{eq:ham-k}) is close to the Dirac
Hamiltonian
$H_D = \frack{\sqrt{3}}{2}(\pm k_x \sigma_x - k_y \sigma_y)$
obtained via a Taylor expansion of the TB Hamiltonian about $K$
\cite{net09}.  The Hamiltonian (\ref{eq:ham-m}) thus captures the
essential features of the graphene multivalley band structure for
both the conduction and valence band near the entire line $K-M-K'$,
so that it provides an alternative approach to \emph{valleytronics}
\cite{two07}.  Unlike $H_D$, the Hamiltonian (\ref{eq:ham-m})
accounts for time reversal symmetry in a natural way.

To discuss edge states we consider graphene ribbons with zigzag
[Fig.~\ref{fig:ribbon}(a)] and armchair edges
[Fig.~\ref{fig:ribbon}(b)].  The electronic states in these ribbons
near energy $E=0$ emerge from the states in 2D graphene which are
highlighted in green in Fig.~\ref{fig:ribbon}(c) and (d).

First we focus on zigzag edges [Fig.~\ref{fig:ribbon}(a)].  We
denote the wave vector for the motion along (perpendicular to) the
direction of the ribbon as $k_\|$ ($k_\perp$).  Ignoring SOC, zigzag
edges give rise to a gapped spectrum around $k_\|=0$ with edge
states appearing in the center of the gap \cite{fuj96}.  These
results are readily rederived by means of Hamiltonian
(\ref{eq:ham-m}), where a suitable coordinate transformation gives
the Hamiltonian
\begin{equation}
  \label{eq:ham-zz}
  H_z(\kk) =
  (1 - \frack{1}{6} k_\perp^2 + \frack{1}{2\sqrt{3}} k_\perp k_\|) \sigma_z
  - (\frack{1}{\sqrt{3}} k_\perp + k_\|) \sigma_y .
\end{equation}
We model the edges as potential steps $b \sigma_z$ with
$b\equiv b_g=0$ inside the ribbon and $b\equiv b_e \ne 0$ outside.
In the end we may consider the limit $|b_e| \rightarrow \infty$ so
that the wave functions vanish at the edges \cite{zho08, son10c}.

\begin{figure}
  \includegraphics[width=0.90\columnwidth]{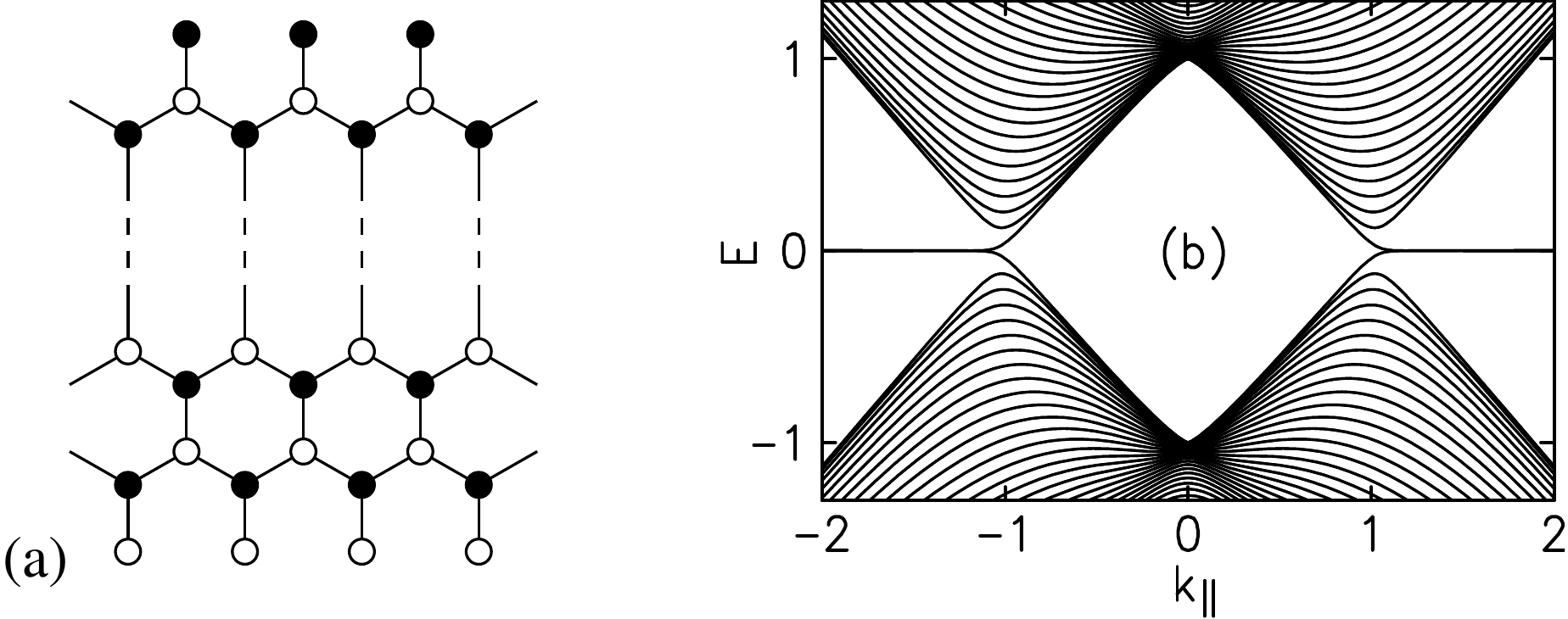}
  \caption{\label{fig:beard}(a) Crystal structure of a graphene
  ribbon with bearded edges.  (b) Band structure
  $E(k_\|)$ of the ribbon in the absence of SOC.  The width of the
  ribbon is $w=40$ and we used $b_e = +10$.}
\end{figure}

The edge states resulting from Eq.\ (\ref{eq:ham-zz}) are shown in
Fig.~\ref{fig:ribbon}(e).  In these calculations, we used the
barrier parameter $b_e = -10$.  Having $b_e < 0$ implies a band
inversion at the graphene edge \cite{vol85}.  A simple confinement
$b_e > 0$ results in the spectrum shown in Fig.~\ref{fig:beard}(b),
where we have the same bulk spectrum as in Fig.~\ref{fig:ribbon}(e),
but the edge states appear for $|k_\|| \gtrsim 1$.  The latter type
of spectrum is obtained in TB calculations for ribbons with
bearded edges \cite{kle94a, ryu02, ber13a}, see
Fig.~\ref{fig:beard}(a).  

We can understand the results in Figs.~\ref{fig:ribbon}(e,g) and
\ref{fig:beard} by looking at the \emph{bulk} eigenstates of the
ribbon.  Figures~\ref{fig:ribbon}(i) and (j) show the squared
magnitude of the up component of the pseudospin $\sigma_z$ [in the
basis of Eq.\ (\ref{eq:ham-zz})] of the lowest $E>0$ bulk
eigenstates as a function of $k_\|$ (with very similar results also
for $b_e > 0$).  Figure~\ref{fig:ribbon}(i) indicates that for small
$|k_\||$ these states are predominantly pseudospin-up.  Yet around
$|k_\|| \simeq 1$ (reflecting the $K$ point of bulk graphene) the
character of these states changes from dominantly spin-up to
spin-down.  The electronic states thus see effective band edges as a
function of $k_\|$ that show a band inversion in reciprocal space
around $|k_\|| \simeq 1$.  This inversion results in robust edge
states for either $|k_\|| \lesssim 1$ [Fig.~\ref{fig:ribbon}(e)] or
$|k_\|| \gtrsim 1$ [Figs.~\ref{fig:beard}(b)], depending on the sign
of $b_e$.  For graphene zigzag ribbons it is, of course, well-known
that different boundary conditions at the edges as in
Figs.~\ref{fig:ribbon}(a) and Figs.~\ref{fig:beard}(a) yield these
different edge states \cite{kle94a, ber13a, fuj96}.  Yet it is,
indeed, a \emph{common} feature of TIs that they show a band
inversion in reciprocal space between the TRIM $\Lambda = 0$ and
$\Lambda = \pi$ of the 1D BZ, so that we have edge states around
either $\Lambda = 0$ or $\Lambda = \pi$.  Choosing appropriate
boundary conditions at the edges of the ribbon thus allows one to
select the location of edge states in the 1D BZ while keeping the
bulk spectrum unaffected.

\begin{figure}
  \includegraphics[width=1.0\columnwidth]{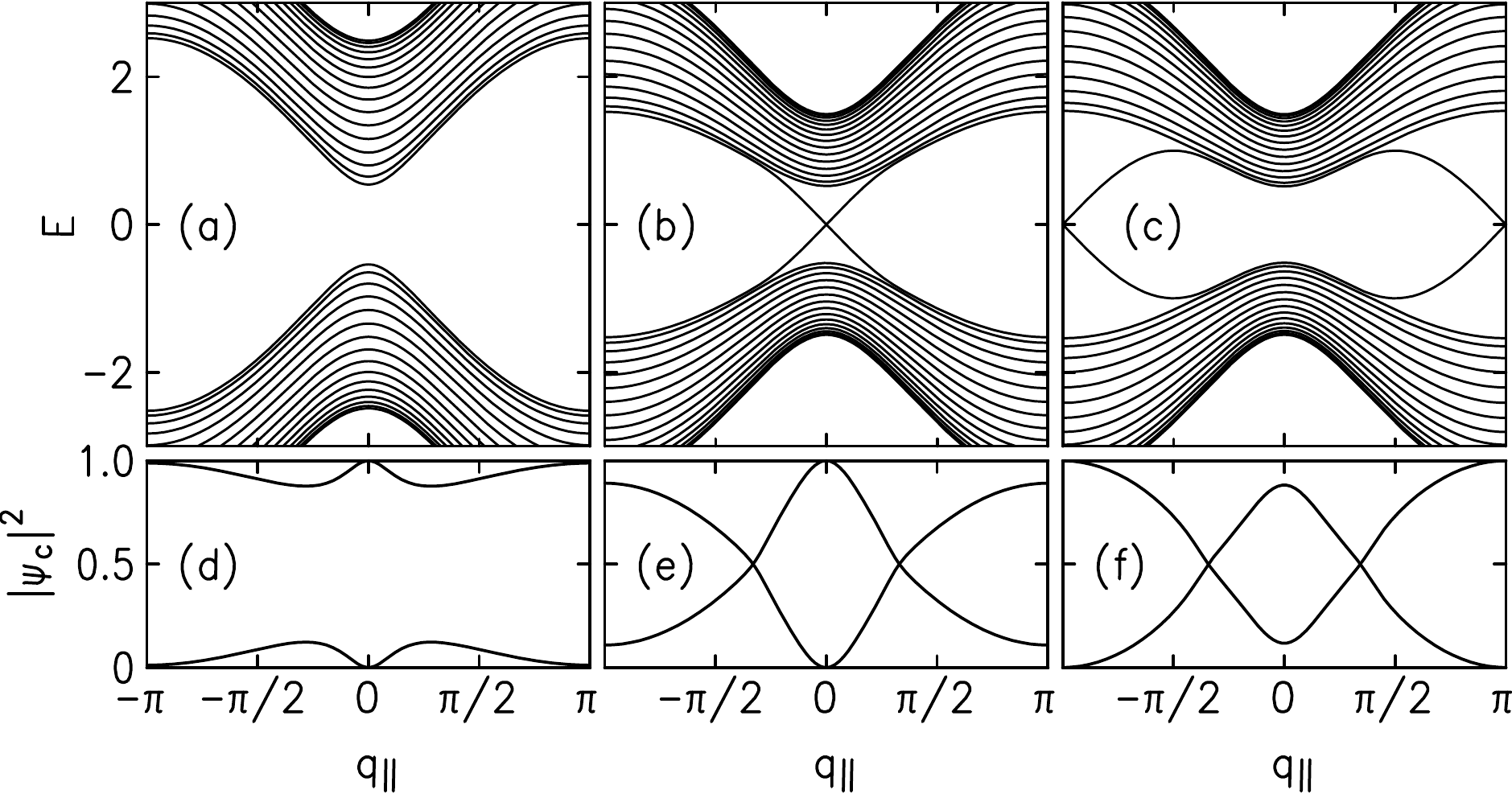}
  \caption{\label{fig:bhz}Band structure $E(q_\|)$ of a BHZ ribbon
  with $a=1$ and $b=0.5$. (a) $m = -0.5$ yields a gapped spectrum
  without edge states. (b) For $m = + 0.5$, the usual boundary
  conditions at the edges of the ribbon \cite{asb16} yield edge
  states near $q_\| = 0$. (c) Dropping one of the basis states
  $s \pm p$ in the outermost layers of the ribbon yields edge states
  near the boundary $q_\| = \pi$ of the BZ. Projection of the lowest
  bulk conduction band states and uppermost bulk valence band states
  on the subspace of positive energies at (d), (e) $q_\| = 0$ and
  (f) $q_\| = \pi$.}
\end{figure}

To illustrate this point, Fig.~\ref{fig:bhz} shows the band
structure of a ribbon using the TB regularization of the BHZ model
based on a square lattice with one $s$ and $p$ orbital per unit cell
\cite{ber06a, asb16}.  The Hamiltonian is
$H = \bigl({h (\qq) \atop 0} {0 \atop h^\ast(-\qq)} \bigr)$ with
$h (\qq) = \vek{d} \cdot \vek{\sigma}$, $d_x = a \sin q_x$,
$d_y = a \sin q_y$, and $d_z = m - 2b (2 - \cos q_x - \cos q_y )$.
In Fig.~\ref{fig:bhz}(a) the mass parameter $m$ is negative,
yielding a trivial regime without edge states.  For $m > 0$ and
using the usual boundary conditions \cite{asb16}, we get conducting
edge states near the center $q_\|=0$ of the 1D BZ
[Fig.~\ref{fig:bhz}(b)].  Alternatively, we may consider the
unitarily equivalent problem with hybridized basis orbitals $s+p$
and $s-p$.  Dropping one of these orbitals in the outermost layers
of the ribbon yields edge states near the boundary $q_\| = \pi$ of
the 1D BZ, while the bulk spectrum remains unchanged
[Fig.~\ref{fig:bhz}(c)].  For the trivial case in
Fig.~\ref{fig:bhz}(a) the bulk eigenstates do not show band
inversion as a function of $q_\|$ [Fig.~\ref{fig:bhz}(d)], whereas
the nontrivial cases in Figs.~\ref{fig:bhz}(b) and (c) show band
inversion [Figs.~\ref{fig:bhz}(e) and (f)].  Similar results also
hold for graphene ribbons with armchair edges \cite{supp}.

We return to the effective Hamiltonian (\ref{eq:ham-m}).  The
numerical calculations presented in this work use a quadrature
method as described in Refs.~\onlinecite{win03, win93a}, which
automatically ensures the proper matching conditions for the
multi-spinor wave function at the edges of the ribbon.  The
numerical results can be confirmed by analytical calculations
similar to those in Refs.~\onlinecite{zho08, son10c}.  In
particular, the limit of hard walls $b_e \rightarrow - \infty$
yields for the edge state at $k_\| = 0$ of a semi-infinite graphene
sheet at $r_\perp \ge 0$
\begin{equation}
  \label{eq:wf-zz}
  \psi_z (r_\perp) = {\Ts \left({1 \atop 1}\right)}
  (e^{-\kappa_+ r_\perp} - e^{-\kappa_- r_\perp}) ,
  \hspace{1em} \kappa_\pm \equiv \sqrt{3}(1\pm i) ,
\end{equation}
and $\psi_z (r_\perp <0) = 0$.  The corresponding eigenenergy is
$E=0$.  The full expressions for finite $b_e$, finite thickness of
the ribbon and finite $k_\|$ are more complicated so that they are
not reproduced here.  Yet such calculations confirm that no edge
states exist around $k_\| = 0$ for $b_e >0$.

For a ribbon with armchair edges [Fig.~\ref{fig:ribbon}(b)] and
neglecting SOC, the effective Hamiltonian becomes
\begin{equation}
  \label{eq:ham-a}
  H_a (\kk) = 
  (1 - \frack{1}{4} k_\perp^2 + \frack{1}{12} k_\|^2 ) \, \sigma_z
  - \frack{2}{\sqrt{3}} k_\| \sigma_y ,
\end{equation}
see Fig.~\ref{fig:ribbon}(d).  The 1D spectrum resulting from
$H_a (\kk)$ is shown in Fig.~\ref{fig:ribbon}(f).  Here we have no
edge states and in the limit of wide ribbons the spectrum becomes
gapless.  These results are in line with the findings in
Ref.~\onlinecite{fuj96}, keeping in mind that the bulk $K$ and $K'$
points are folded to the point $q_\| = 0$ of the 1D BZ for armchair
ribbons.

Previously Brey and Fertig \cite{bre06a} used the graphene Dirac
Hamiltonian $H_D$ to obtain the edge states of 1D ribbons emerging
from the states near the $K$ point of 2D graphene \cite{boundary},
i.e., their model gives the edge states for wave vectors close to
the valence band maximum and conduction band minimum in
Figs.~\ref{fig:ribbon}(e) and (f).  The present approach is
different from this earlier work as it yields the edge states in the
entire region in between the points $K$ and $K'$ of the bulk band
structure including the robust crossing at $k_\| = 0$, consistent
with the TB description \cite{fuj96}.  Effective Hamiltonians based
on a Taylor expansion of the band structure are often low-energy
Hamiltonians that are valid only in the vicinity of the expansion
point \cite{bas88, win03}.  Yet this is not an inherent constraint
\cite{car66}.

Next we discuss the effect of SOC.  First we consider
$\lambda_r = 0$.  The intrinsic SOC $\propto \lambda_i$ opens a gap
$16 \lambda_i \sqrt{1 - 16\lambda_i^2} \approx 16 \lambda_i$ in the
bulk spectrum of the Hamiltonian (\ref{eq:ham-m}).  In the TB model
this gap becomes $6\sqrt{3}\lambda_i$ \cite{kan05a}.  For
$\lambda_i \ne 0$ the edge states in a zigzag or armchair ribbon
remain two-fold degenerate at $k_\| = 0$ which reflects the fact
that these states originate from the time-reversal invariant $M$
point of the graphene BZ. This aspect is thus readily captured by
the Hamiltonian (\ref{eq:ham-m}), as demonstrated in
Figs.~\ref{fig:ribbon}(g) and (h).

\begin{figure}[t]
  \includegraphics[width=0.85\columnwidth]{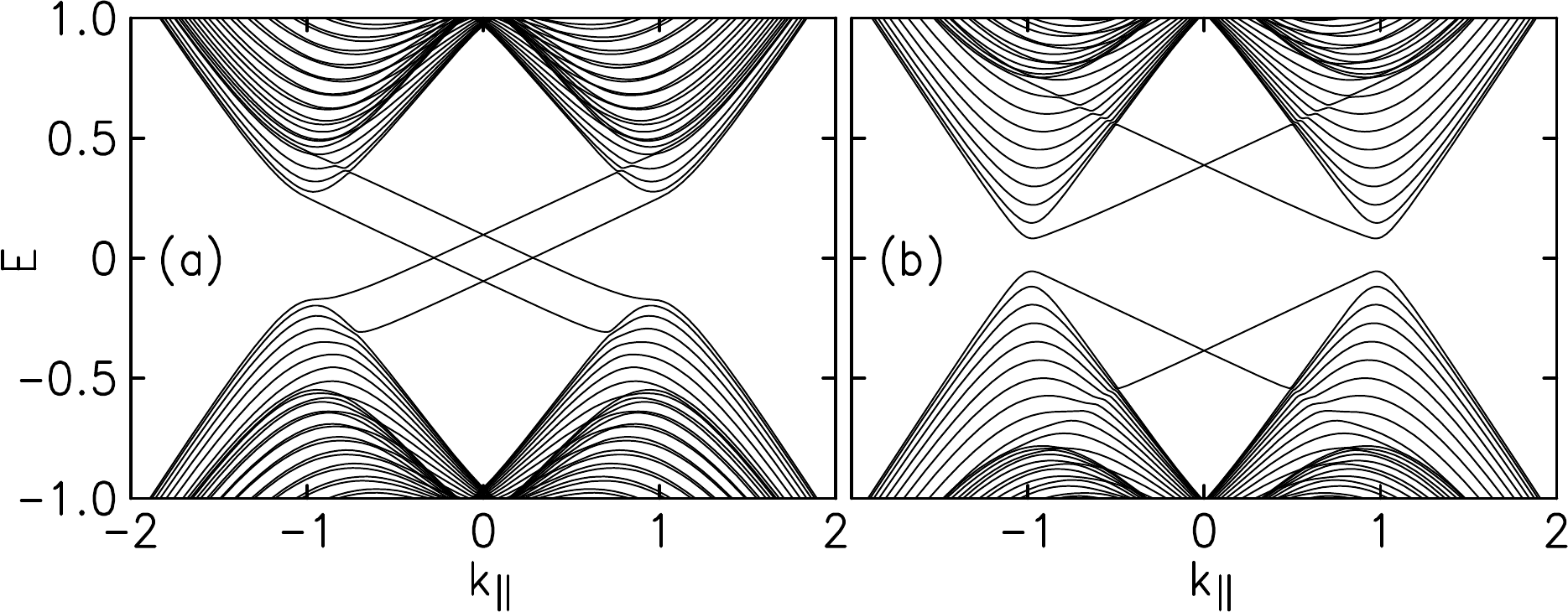}
  \caption{\label{fig:kane}Band structure $E(k_\|)$
  of zizag ribbons for $\lambda_i = 0.09$ and $\lambda_r = 0.05$.
  The sublattice staggering is (a) $\lambda_v = 0.1$ and (b)
  $\lambda_v = 0.4$.  The width of the ribbon is $w=40$ and we used
  $b_e = -10$.  Compare Fig.~1 of Ref.~\onlinecite{kan05}.}
\end{figure}

Both the intrinsic SOC $\propto \lambda_i$ and the staggering
$\propto \lambda_v$ open a gap in the bulk spectrum of the
Hamiltonian (\ref{eq:ham-m}).  Yet it follows immediately from Eq.\
(\ref{eq:ham-k}) that the gap closes for
$\lambda_v = \pm 8 \lambda_i$.  Consistent with the TB results in
Ref.~\onlinecite{kan05} this set of parameters describes the phase
boundary between the topologically trivial regime with an even
number of edge states and the nontrivial regime with an odd number
of edge states crossing the bulk gap.  Similarly, Rashba SOC
$\propto \lambda_r$ induces such a phase transition when it competes
with the intrinsic SOC.  Figure~\ref{fig:kane} illustrates this
point for zigzag ribbons.  Here the staggering $\lambda_v = 0.1$
[Fig.~\ref{fig:kane}(a)] gives rise to edge states crossing the gap,
whereas $\lambda_v = 0.4$ [Fig.~\ref{fig:kane}(b)] results in an
ordinary insulator.  These calculations are in very good agreement
with the TB results in Fig.~1 of Ref.~\onlinecite{kan05}.

Finally we comment on the general robustness of the edge states,
which is a major aspect motivating the interest in topological
insulators \cite{has10, qi11}.  It was pointed out in
Ref.~\onlinecite{kan05a} that the edge states at $\Lambda \pm k_\|$
form Kramers doublets so that elastic backscattering from a weak
random potential preserving time reversal symmetry is forbidden.
This argument applies also to the effective Hamiltonian
(\ref{eq:ham-m}).  Yet if the electron states are modeled by means
of the low-energy Dirac Hamiltonian $H_D$, the two time-reversed
valleys at $K$ and $K'$ are described via a discrete valley
pseudospin degree of freedom, that makes it difficult to incorporate
intervalley scattering in a general way.  For the Hamiltonian
(\ref{eq:ham-m}) pairs of time-reversed states are connected by
continuous paths in the Hilbert space of this Hamiltonian so that it
is well-suited to incorporate intervalley scattering, though a
detailed study of this point is beyond the scope of the present
work.

In conclusion, the effective Hamiltonian (\ref{eq:ham-m}) based on
an expansion of the graphene TB Hamiltonian about the time-reversal
invariant $M$ point provides an accurate description of the
topologically protected edge states in graphene, although the
concept of a BZ is not part of such an effective model.  Similar
effective Hamiltonians can be derived via a Taylor expansion of,
e.g., the graphene TB Hamiltonian about the BZ center $q = 0$ or the
TB-regularized BHZ Hamiltonian about the TRIM $\qq = (\pi,0)$
[suitable for a description of the ``inverted'' protected edge
states in Fig.~\ref{fig:bhz}(c)].  We may expect that similar
Hamiltonians exist also for other topologically protected systems in
both 2D and 3D.  Our work may inspire further research on necessary
conditions for the formation of protected edge states and robust
level degeneracies.  For example, numerical studies based on the
effective $8\times 8$ Kane Hamiltonian have shown that a 2D Dirac
semimetal with robust level crossings can be realized in HgTe-CdTe
quantum wells when the well thickness is varied \cite{win12a}.
Also, the effective models proposed here open an avenue for studying
these systems under perturbations such as homogeneous or
inhomogeneous \cite{hal88} magnetic and electric fields or strain
which may break the periodicity of the ideal crystal structure
\cite{bir74} so that it is more difficult to incorporate such
effects in atomistic calculations.
RW appreciates stimulating discussions with C.~S.\ Chu, C.~L.\ Kane,
A.~P\'alyi, and U.~Z\"ulicke.  This work was supported by the NSF
under grant No.\ DMR-1310199.  Work at Argonne was supported by DOE
BES under Contract No.\ DE-AC02-06CH11357.

\onecolumngrid
\cleardoublepage

\setcounter{equation}{0}
\renewcommand{\theequation}{S\arabic{equation}}
\setcounter{figure}{0}
\renewcommand{\thefigure}{S\arabic{figure}}
\thispagestyle{empty}

\begin{center}
  {\large\bf Supplemental Material:}\\[0.5ex]
  {\large\bf Effective Hamiltonian for protected edge states in graphene}\\[2ex]
  H.~Deshpande and R.~Winkler
\end{center}

\vspace{2ex}
\twocolumngrid

\begin{figure}
  \includegraphics[width=\columnwidth]{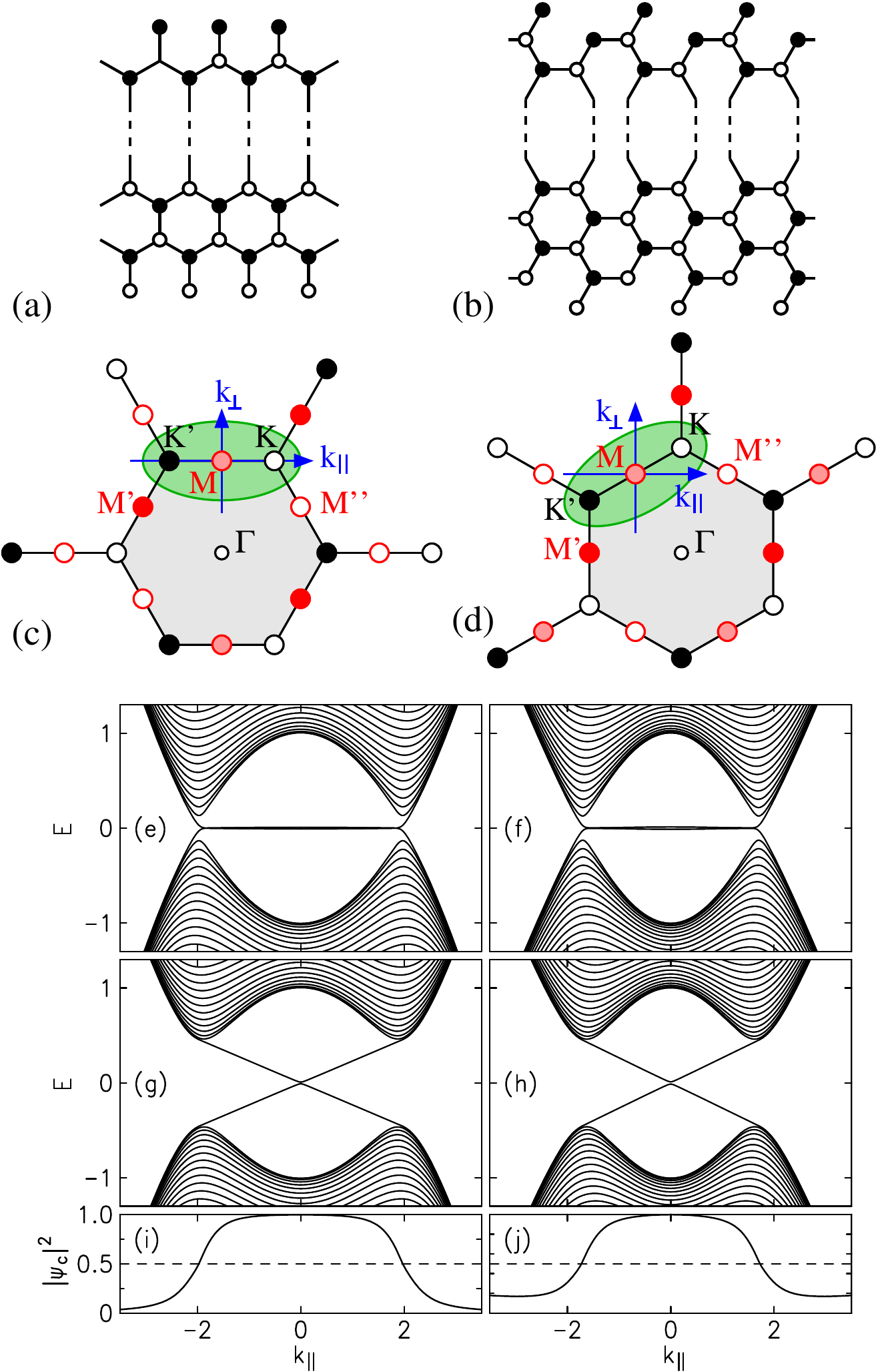}
  \caption{\label{fig:bd}(Color online) Crystal structure of a
  graphene ribbon with (a) bearded zigzag and (b) bearded armchair
  edges.  Bulk BZ of graphene corresponding to (c) bearded zigzag
  and (d) bearded armchair ribbons.  The region of the BZ captured
  by the Hamiltonian (\ref{eq:ham-m}) is marked in green.  Band
  structure $E(k_\|)$ of (e) bearded zigzag and (f) bearded armchair
  ribbons in the absence of SOC and (g), (h) for $\lambda_i = 0.2$.
  (i), (j) Squared magnitude of the up component of pseudospin
  $\sigma_z$ of the lowest $E>0$ bulk eigenstates.  The width of the
  ribbons is $w=40$ and we used $b_e = -10$.}
\end{figure}

We show here that the protected edge states in graphene ribbons with
bearded zigzag [Fig.~\ref{fig:bd}(a)] and bearded amrchair edges
[Fig.~\ref{fig:bd}(b)] can be analyzed in complete analogy with
Fig.~\ref{fig:ribbon}.  The electronic states in these ribbons near
energy $E=0$ emerge from the states in the 2D BZ of graphene which
are highlighted in green in Fig.~\ref{fig:bd}(c) and (d).  First we
discuss these ribbons in the absence of SOC.  For bearded zigzag
edges, the effective Hamiltonian becomes
\begin{equation}
  H_{bz} (\kk) =
    \bigl(1 - \frack{1}{4} k_\|^2 + \frack{1}{12} k_\perp^2 \bigr) \sigma_z
    + \frack{2}{\sqrt{3}} k_\perp \sigma_y .
\end{equation}
The resulting dispersion is shown in Fig.~\ref{fig:bd}(e).  For
bearded armchair edges, the effective Hamiltonian becomes
\begin{equation}
  H_{ba} (\kk) =
  (1 - \frack{1}{6} k_\|^2 + \frack{1}{2\sqrt{3}} k_\| k_\perp) \sigma_z
  - (\frack{1}{\sqrt{3}} k_\| + k_\perp) \sigma_y .
\end{equation}
The resulting dispersion is shown in Fig.~\ref{fig:bd}(f).  Similiar
to Figs.~\ref{fig:ribbon}(g) and (h), intrinsic SOC opens a gap in
the bulk band structure of bearded ribbons.  Using $b_e <0$ we get
the protected edge states shown in Figs.~\ref{fig:bd}(g) and (h).
Similar to Figs.~\ref{fig:ribbon}(e)-(h), the results in
Figs.~\ref{fig:bd}(e)-(h) are in very good agreement with TB
calculations for the ribbon geometries in Figs.~\ref{fig:bd}(a) and
(b).  We note that similar to Figs.~\ref{fig:bhz}(b) and (c) the
bulk spectra in Figs.~\ref{fig:ribbon}(e), (g) and
Figs.~\ref{fig:bd}(e), (g) are the same, independent of the
spectra of the edge states.

Figures~\ref{fig:beard}(b) (obtained for $b_e > 0$)
and~\ref{fig:bd}(e) (obtained for $b_e < 0$) refer to the same
bearded ribbon shown in Fig.~\ref{fig:bd}(a).  For this ribbon, the
protected edge states disperse symmetrically about the TRIM
$\Lambda = 0$ of the 1D BZ.  An essential difference between the
effective Hamiltonians underlying these figures thus lies in the
fact that only the expansion point $\vek{M}$ in Fig.~\ref{fig:bd}(c)
is mapped onto $\Lambda = 0$ of the 1D BZ, thus yielding a robust
model for the protected edge states of bearded zigzag ribbons, as
demonstrated by Fig.~\ref{fig:bd}(g).

More generally, a 2D TI has four TRIM $\Lambda$ that are mapped
pairwise on the two 1D-TRIM $\Lambda = 0$ and $\Lambda = \pi$ of the
corresponding TI ribbon.  The edge states may disperse about either
$\Lambda = 0$ or $\Lambda = \pi$, and we need to choose the
expansion point of the effective Hamiltonian accordingly for a
robust description of the protected edge states.


\begin{thebibliography}{10}

\bibitem{kan05}
C.~L. Kane and E.~J. Mele, Phys.\ Rev.\ Lett. \textbf{95}, 146802 (2005).

\bibitem{kan05a}
C.~L. Kane and E.~J. Mele, Phys.\ Rev.\ Lett. \textbf{95}, 226801 (2005).

\bibitem{has10}
M.~Z. Hasan and C.~L. Kane, Rev.\ Mod.\ Phys. \textbf{82}, 3045 (2010).

\bibitem{qi11}
X.-L. Qi and S.-C. Zhang, Rev.\ Mod.\ Phys. \textbf{83}, 1057 (2011).

\bibitem{koe07}
M.~K\"onig, S.~Wiedmann, C.~Br\"une, A.~Roth, H.~Buhmann, L.~W. Molenkamp,
  X.-L. Qi, and S.-C. Zhang, Science \textbf{318}, 766 (2007).

\bibitem{ber06a}
B.~A. Bernevig, T.~L. Hughes, and S.-C. Zhang, Science \textbf{314}, 1757
  (2006).

\bibitem{liu08}
C.~Liu, T.~L. Hughes, X.-L. Qi, K.~Wang, and S.-C. Zhang, Phys.\ Rev.\ Lett.
  \textbf{100}, 236601 (2008).

\bibitem{zho08}
B.~Zhou, H.-Z. Lu, R.-L. Chu, S.-Q. Shen, and Q.~Niu, Phys.\ Rev.\ Lett.
  \textbf{101}, 246807 (2008).

\bibitem{son10c}
E.~B. Sonin, Phys.\ Rev.~B \textbf{82}, 113307 (2010).

\bibitem{effective}
Effective models using a Taylor expansion of the band structure are based on
  the assumption that the resulting model can describe the states near the
  Fermi energy. They would generally fail to describe both ``topological'' and
  ``trivial'' (nontopological) properties of a material if we have bands in
  other parts of the Brillouin zone near the Fermi energy.

\bibitem{wal47}
P.~R. Wallace, Phys.\ Rev. \textbf{71}, 622 (1947).

\bibitem{hal88}
F.~D.~M. Haldane, Phys.\ Rev.\ Lett. \textbf{61}, 2015 (1988).

\bibitem{net09}
A.~H. {Castro Neto}, F.~Guinea, N.~M.~R. Peres, K.~S. Novoselov, and A.~K.
  Geim, Rev.\ Mod.\ Phys. \textbf{81}, 109 (2009).

\bibitem{prefac}
To match the conventions in Ref.~\onlinecite{win10a}, the intrinsic SOC term in
  Eq.\ (\ref{eq:ham-m}) must be multiplied by a factor $-1/2\sqrt{3}$ and the
  Rashba term by $-1/\sqrt{3}$.

\bibitem{win10a}
R.~Winkler and U.~Z\"ulicke, Phys.\ Rev.~B \textbf{82}, 245313 (2010).

\bibitem{nov05a}
K.~S. Novoselov, A.~K. Geim, S.~V. Morozov, D.~Jiang, M.~I. Katsnelson, I.~V.
  Grigorieva, S.~V. Dubonos, and A.~A. Firsov, Nature \textbf{438}, 197 (2005).

\bibitem{two07}
J.~Tworzyd{\l}o, I.~Snyman, A.~R. Akhmerov, and C.~W.~J. Beenakker, Phys.\
  Rev.~B \textbf{76}, 035411 (2007).

\bibitem{fuj96}
M.~Fujita, K.~Wakabayashi, K.~Nakada, and K.~Kusakabe, J.~Phys.\ Soc.\ Jpn.
  \textbf{65}, 1920 (1996).

\bibitem{vol85}
B.~A. Volkov and O.~A. Pankratov, JETP Lett. \textbf{42}, 178 (1985).

\bibitem{kle94a}
D.~J. Klein, Chem. Phys. Lett. \textbf{217}, 261 (1994).

\bibitem{ryu02}
S.~Ryu and Y.~Hatsugai, Phys.\ Rev.\ Lett. \textbf{89}, 077002 (2002).

\bibitem{ber13a}
B.~A. Bernevig, \emph{Topological Insulators and Topological Superconductors}
  (Princeton University Press, Princeton, NJ, 2013).

\bibitem{asb16}
J.~K. Asb\'oth, L.~Oroszl\'any, and A.~P\'alyi, \emph{A Short Course on
  Topological Insulators} (Springer, Cham, 2016).

\bibitem{supp}
See supplemental material.

\bibitem{win03}
R.~Winkler, \emph{Spin-Orbit Coupling Effects in Two-Dimen\-sion\-al Electron
  and Hole Systems} (Springer, Berlin, 2003).

\bibitem{win93a}
R.~Winkler and U.~R\"ossler, Phys.\ Rev.~B \textbf{48}, 8918 (1993).

\bibitem{bre06a}
L.~Brey and H.~A. Fertig, Phys.\ Rev.~B \textbf{73}, 235411 (2006).

\bibitem{boundary}
Similar to the present work, boundary conditions played also an important role
  in Ref.~\onlinecite{bre06a}. Yet the nature of these boundary conditions was
  very different from the boundary conditions used here.
  Reference~\onlinecite{bre06a} treated the edges as hard walls. As $H_D$ is
  linear in momentum we cannot require that both spinor components vanish
  simultaneously at the boundaries.

\bibitem{bas88}
G.~Bastard, \emph{Wave Mechanics Applied to Semiconductor Heterostructures}
  (Les Editions de Physique, Les Ulis, 1988).

\bibitem{car66}
M.~Cardona and F.~H. Pollak, Phys.\ Rev. \textbf{142}, 530 (1966).

\bibitem{win12a}
R.~Winkler, L.~Y. Wang, Y.~H. Lin, and C.~S. Chu, Solid State Commun.
  \textbf{152}, 2096 (2012).

\bibitem{bir74}
G.~L. Bir and G.~E. Pikus, \emph{Symmetry and Strain-Induced Effects in
  Semiconductors} (Wiley, New York, 1974).

\end{thebibliography}
\end{document}